\documentclass[%
reprint,
nofootinbib,
 amsmath,amssymb,
 aps,
 prd,
floatfix,
]{revtex4-2}

\usepackage{graphicx}
\usepackage{dcolumn}
\usepackage{adjustbox}
\usepackage{bm}
\usepackage{aas_macros}
\usepackage{xcolor}
\begin{document}

\preprint{APS/123-QED}

\title{Cosmological Constraints from Weak Lensing Peaks: Can Halo Models Accurately Predict Peak Counts?}
\author{Alina Sabyr$^{1}$}
\email{as6131@columbia.edu}
\author{Zolt\'an Haiman$^{1}$}%
\author{Jos\'{e} Manuel Zorrilla Matilla$^{2}$}
\author{Tianhuan Lu$^{1}$}
\affiliation{$^{1}$Department of Astronomy, Columbia University, New York, NY 10027, USA}
\affiliation{$^{2}$Department of Astrophysical Sciences, Peyton Hall, Princeton University, Princeton, New Jersey 0854, USA}
\date{\today}

\begin{abstract}
In order to extract full cosmological information from next-generation large and high-precision weak lensing (WL) surveys (e.g. Euclid, Roman, LSST), higher-order statistics that probe the small-scale, non-linear regime of large scale structure (LSS) need to be utilized. WL peak counts, which trace overdensities in the cosmic web, are one promising and simple statistic for constraining cosmological parameters.
The physical origin of WL peaks have previously been linked to dark matter halos along the line of sight and this peak-halo connection has been used to develop various semi-analytic halo-based models for predicting peak counts. 
Here, we study the origin of WL peaks and the effectiveness of halo-based models for WL peak counts using a suite of ray-tracing N-body simulations.  We compare WL peaks in convergence maps from the full simulations to those in maps created from only particles associated with halos -- the latter playing the role of a ``perfect'' halo model. We find that while halo-only contributions are able to replicate peak counts qualitatively well, halos do not explain all WL peaks. Halos particularly underpredict negative peaks, which are associated with local overdensities in large-scale underdense regions along the line of sight.  In addition, neglecting non-halo contributions to peaks counts leads to a significant bias on the parameters ($\Omega_{\rm m}$, $\sigma_{8}$) for surveys larger than $\gtrapprox$ 100 deg$^{2}$. We conclude that other elements of the cosmic web, outside and far away from dark matter halos, need to be incorporated into models of WL peaks in order to infer unbiased cosmological constraints.
\end{abstract}

\maketitle

\section{Introduction}

Weak gravitational lensing (WL) by large scale structure (LSS) is one of the most promising cosmological probes of both dark matter (DM) and dark energy (DE). It traces the mass distribution in the universe and is sensitive both to the expansion history and growth of structure (e.g.~\cite{Bartelmann2001,Refregier2003,HoekstraJain2008,Kilbinger2015}). 
Previous and ongoing galaxy surveys, such as the Canada-France-Hawaii Telescope Lensing Survey (CFHTLenS; e.g.~\cite{Kilbinger2013}), the Kilo-Degree Survey (KiDS; e.g.~\cite{Hildebrandt2017}), the Dark Energy Survey (DES; e.g.~\cite{DES2018,DES2021}) and the Hyper SuprimeCam (HSC; e.g.~\cite{Hikage2019}) have already shown that WL is able to provide competitive constraints on cosmological parameters. Upcoming next-generation surveys such as the Vera C. Rubin Observatory's 10-year Legacy Survey of Space and Time (LSST)\footnote{https://www.lsst.org/}, the Nancy Grace Roman Space Telescope\footnote{https://roman.gsfc.nasa.gov/}, Euclid\footnote{https://sci.esa.int/web/euclid}, and the Square Kilometre Array (SKA)\footnote{https://www.skatelescope.org/} will deliver WL data covering larger areas at higher resolution.

On large scales, the matter distribution is linear or quasi-linear and can be described by second-order statistics, such as the two-point correlation function or its Fourier transform, the power spectrum. On small scales, however, gravitational collapse causes the WL signal to be dominated by non-linear mass fluctuations. In order to extract the full cosmological information content from these new datasets that will probe deeper into the non-linear regime, higher-order statistics need to be utilized. Several different observables have been proposed to obtain such non-Gaussian information from WL. Among these are, for example, 
the three-point correlation function~\cite{Takada2003}
or its Fourier transform, the bispectrum~\cite{TakadaJain2004, DodelsonZhang2005}, Minkowski functionals~\cite{Kratochvil2012, Munshi+2012, ShirasakiYoshida2014, Petri2015} and lensing peaks~\cite{Jain2000, DietrichHartlap2010, Kratochvil2010}.

Lensing peaks, which are defined as local maxima on WL maps, are a simple statistic that can be easily measured over large datasets. Peaks in (e.g.) convergence maps correspond to overdensities in projected LSS. Hence, their physical origin has previously been studied both theoretically~\cite{Yang2011} and in observational CFHTLenS data~\cite{Liu2016}. Ref.~\cite{Yang2011} showed that high-significance peaks, whose signal-to-noise ratio (S/N) exceeds 3.5, are typically due to a single massive halo, while lower-significance peaks have been linked to constellations of multiple, less massive halos along the line of sight (LOS). The latter were found to be particularly sensitive to cosmological parameters. Recently, Ref. \cite{Wei2018} studied correlations between WL peaks and halos, and found that in WL maps including galaxy shape noise, $\approx$60\% of high peaks (S/N $\geq 5$) and $\approx$20\% of lower peaks ($3 \leq S/N < 5$) are associated with halos along the LOS. 

Numerical simulations used for predicting peak counts can be computationally expensive, which has motivated the development of semi-analytic and stochastic models, based on the physical nature of WL peaks. For example,
Ref.~\cite{Maturi2010} used Gaussian random fields to analytically predict peak counts caused by chance superpositions of LSS and noise (although note that they use a different definition of a ``peak'', based on genus statistics).  They found that peaks with S/N up to 3$-$5 are dominated by these chance superpositions, rather than by dark matter halos. 

Ref. \cite{Fan2010} put forward a model incorporating both a mass function of Navarro--Frenk--White (NFW) dark matter halos and galaxy shape noise to predict the total number of peak counts. More recently, Ref. \cite{Yuan2018} extended this model by using random Gaussian fields to incorporate LSS effects, which become important for larger and deeper sky surveys. Ref. \cite{Giocoli2017} presented a fast model for simulating weak lensing maps using a halo model formalism, which takes as an input the linear power spectrum of the cosmological model and a halo catalog. Ref.~\cite{Giocoli2018} have checked the accuracy of this model for peak counts and found that it works well (to within 10\%) but not at a percent level precision for all peaks.

Similarly, the publicly-available code CAMELUS~\cite{Lin2015a,Lin2015b} is a fast stochastic halo-based simulation that avoids the high computational cost associated with evolving N-body simulations. Instead, CAMELUS samples a mass function to populate a volume with a set of halos, all of which are assigned a density profile and placed randomly. Relying on fast halo-based models can be advantageous as it offers a simpler way to simulate maps for larger upcoming surveys. CAMELUS was shown to predict the abundance of most WL peaks accurately \cite{Lin2015a}. However, given the stringent requirements in large WL surveys, CAMELUS cannot be used, as is, for cosmological parameter estimates. 

Ref.~\cite{Zorrilla2016} (hereafter ZM16) extended the comparison between CAMELUS and N-body simulations to a suite of 162 different cosmologies. They found that, depending on cosmology, there are significant ($\sim 20\%)$ differences, especially in the number of low-S/N and negative peaks, even in the presence of galaxy shape noise.   Furthermore, the (co)variances in the peak counts predicted by CAMELUS are much lower, and much less sensitive to the background cosmology, than peaks in N-body simulations.  Finally, CAMELUS was found to rely primarily on high-significance peaks for cosmological parameter predictions, whereas lower-amplitude peaks contribute most of the constraints in the N-body simulations.

These remaining discrepancies between lensing peaks in CAMELUS and N-body simulations could be attributed to several effects. (1) {\it Non-halo contributions} are present in N-body simulations but excluded in CAMELUS.(2) Likewise, CAMELUS neglects the {\it spatial clustering} of halos. (3) CAMELUS halos are spherically symmetric and are assigned {\it NFW profiles}, whereas N-body halos are asymmetric and their profiles do not match NFW exactly. (4) Finally, CAMELUS  assigns a {\it fixed concentration parameter} to all halos of a given mass and redshift, whereas N-body halos, at a fixed mass and redshift, have a range of shapes and profiles.
In this study, we follow up on ZM16 to address these effects.  We are motivated by the fact that the physical origin of low-significance WL peaks remains unclear, and by the need for fast and simple predictions for peak counts and (co)variances, suitable for fitting data obtained in large forthcoming surveys. In this work, we examine whether halos alone can explain WL peaks. We do so by comparing the peak statistics obtained in WL maps in ray-traced full N-body simulations to peak statistics obtained in ray-traced maps containing contributions only from halos.  The latter represents the limiting case of a ``perfect'' halo model.

The rest of this paper is organized as follows. In \S~II, we describe our suite of N-body simulations, our raytracing pipeline used for creating WL maps, how we created halo-only maps, and the statistical inference methods we use to constrain cosmological parameters. In \S~III, we present our results, including peak histograms for various cosmologies, contour plots and correlation matrices. In \S~IV, we discuss our main findings and limitations. We summarize our work and its implications in \S~V.  

\section{Methodology}\label{methodology}

\subsection{N-body simulations.}

\begin{figure}[th]
\includegraphics[width=\columnwidth]{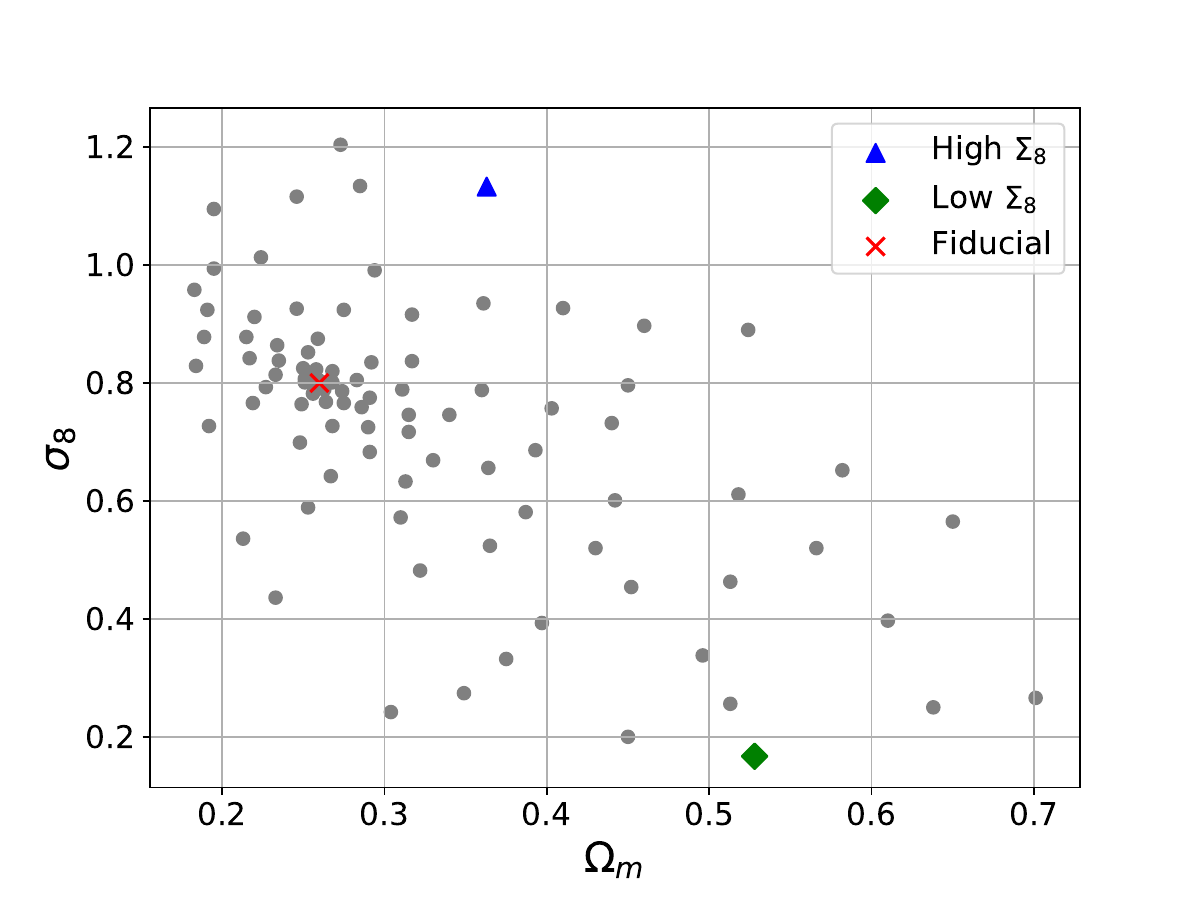}
\caption{\label{fig:grid} 97 cosmologies used for ray-tracing, shown in the region of parameter space used for interpolation. The fiducial model, as well as examples of a low- and a high-$\Sigma_{8}$ cosmology that we use for peak count histograms in Fig. \ref{fig:peakhist} are marked.}
\end{figure}

In this work, we use 97 dark matter (DM)-only N-body simulations in the $\Omega_{\rm m}-\sigma_{8}$ parameter space (Fig.~\ref{fig:grid}). The simulations were run by first computing an initial power spectrum for a set of $\Omega_{\rm m}-\sigma_{8}$ cosmological parameters using the Boltzmann code CAMB \cite{CAMB}, followed by the generation of a single initial displacement field of particles for each cosmology, using the Zeldovich-approximation code NGEN-IC \cite{NGENIC}. Finally, each (240 $h^{-1}$ Mpc)$^{3}$ comoving volume with $512^{3}$ DM particles was evolved from its initial conditions using Gadget-2 \cite{Springel2005}. We saved a total of 59 snapshots at uniform co-moving distances up to a redshift of $\approx3$ to build past light-cones. The resulting mass resolution is $\approx 10^{10} h^{-1} \mathrm{\rm M}_\odot$. For our fiducial $\Lambda$CDM model, we adopt a matter density $\Omega_{\rm m}=0.260$, dark energy density $\Omega_{\rm DE}=1-\Omega_{\rm m}=0.740$ and matter fluctuation amplitude $\sigma_{8}=0.800$ (see Table ~\ref{tab:table1}). 

\begin{table}[ht]
\caption{\label{tab:table1}%
Cosmological parameters of the fiducial model.
}
\begin{ruledtabular}
\begin{tabular}{lcc}
Parameter & Symbol & Value
\\
\colrule
\\
Matter density &$\Omega_{\rm m}$ & 0.260 \\
Dark energy density&$\Omega_{\rm DE}$ & $1.0-\Omega_{\rm m}$\\
Amplitude of fluctuations at 8 $h^{-1}$ Mpc &$\sigma_{8}$&0.800\\
Dimensionless Hubble constant & $h_0$ & 0.72\\
Dark energy equation of state parameter & $w$&$-1.0$\\
\end{tabular}
\end{ruledtabular}
\end{table}

To build weak lensing convergence maps, we use the \texttt{Lenstools} Python package \cite{Petri2016}. First, we slice each N-body snapshot along its three coordinate axes into $80$~$h^{-1}$ Mpc thick lensing planes, and compute their two-dimensional gravitational potential on a $4,096\times4,096$ regular grid. We apply random shifts and rotations to the planes and stack them between the observer and the source galaxies, which are placed at the fixed redshift of $z_{\rm s}=1$. We then follow a standard multi-plane raytracing algorithm by tracing the light rays from the observer to the source galaxies and calculating the corresponding lensing deflections and distortions at each plane to compute $1,024\times1,024$ pixel convergence maps (covering $3.5\times3.5$ deg$^2$ at $z_{\rm s}=1$). Applying random shifts and rotations to the potential planes allows us to recycle a single N-body simulation to generate 500 pseudo-independent convergence maps in each cosmology. A more detailed description of our WL map generation pipeline can be found in \cite{Petri2016}. 
The intrinsic shapes of lensed source galaxies are unknown. Therefore, to account for the contribution of the galaxies' intrinsic ellipticity to the lensing signal derived from shape measurements, we add white noise to our maps by drawing from a zero-mean Gaussian distribution with standard deviation
\begin{equation}
    \sigma_{\rm pix}=\sqrt{\frac{\sigma_{\epsilon}^{2}}{2n_{\rm g}A_{\rm pix}}}
\end{equation}
where we adopt an intrinsic r.m.s. ellipticity of $\sigma_{\epsilon}$=0.4, galaxy surface density $n_{\rm g}=25$ arcmin$^{-2}$ and pixel area $A_{\rm pix} = $ (3.5$^{2}$ deg$^{2}$/1024$^{2}$).

We smooth both the simulated noiseless maps and the Gaussian shape noise maps using a Gaussian filter with standard deviation $\Theta_{\rm G}=1$ arcmin before combining them. Unless otherwise stated, our results are based on WL maps that have been smoothed and include noise.

Throughout this paper, we quote our WL peak values in terms of their significance level or S/N. The conversion factor from convergence to S/N is the average standard deviation across the 500 convergence maps in the fiducial model, $\sigma_{\kappa}=0.02$. As a reference, the r.m.s. value for full N-body, halo-only and non-halo maps (see below) are $\sigma_{\kappa}=0.01988$, $\sigma_{\kappa}=0.01926$, and $\sigma_{\kappa}=0.01625$, respectively.

\subsection{Convergence Maps from Halos and Non-Halo Material}\label{halomaps}

To investigate the contribution of DM halos to peak counts, we first identify halos in the outputs of the N-body simulations using the \texttt{Rockstar} halo-finder~\cite{Behroozi2013}. \texttt{Rockstar} uses an algorithm based on adaptive hierarchical refinement of friends-of-friends (FOF) groups in six phase-space dimensions and one time dimension. The parameters used are a force resolution of 0.009 (to match the resolution of our Gadget-2 simulations) and a FOF linking length of 0.28 (\texttt{Rockstar}'s default value). Some information from \texttt{Rockstar} outputs, such as the biggest and smallest halos identified at redshifts of $z=1$ and $z=0$, and the number of particles associated with all halos at those redshifts, are listed in Table~\ref{tab:table2}.

\begin{table}[t]
\caption{\label{tab:table2}%
Halo information based on \texttt{Rockstar}'s outputs for N-body snapshots at two redshifts. Roughly half (third) of the particles present in the N-body simulations at $z=0$ ($z=1$) are identified by \texttt{Rockstar} as belonging to a halo.
}
\begin{ruledtabular}
\begin{tabular}{lcc}
&
\textit{z} = 0 & \textit{z} = 1\\
\colrule
\\
Number of Halo Particles & 64508061 & 41679024\\
& ($\approx48\%$) & ($\approx 31\%$) \\
\\
Most Massive Halo\\
$M_{\rm vir}$ [${\rm M}_{\odot}$/$h$] & $1.0821\times10^{15}$ & $3.0468\times10^{14}$ \\
$R_{\rm vir}$ [kpc/$h$]  & 2135.288 & 1696.498 \\
\\
Least Massive Halo\\

$M_{\rm vir}$ [${\rm M}_{\odot}$/$h$] & $1.4865\times10^{10}$ & $1.48650\times10^{10}$ \\
$R_{\rm vir}$ [kpc/$h$] & 51.139 & 61.991 \\
\end{tabular}
\end{ruledtabular}
\end{table}

\begin{figure*}[!ht]
\includegraphics[width=\textwidth]{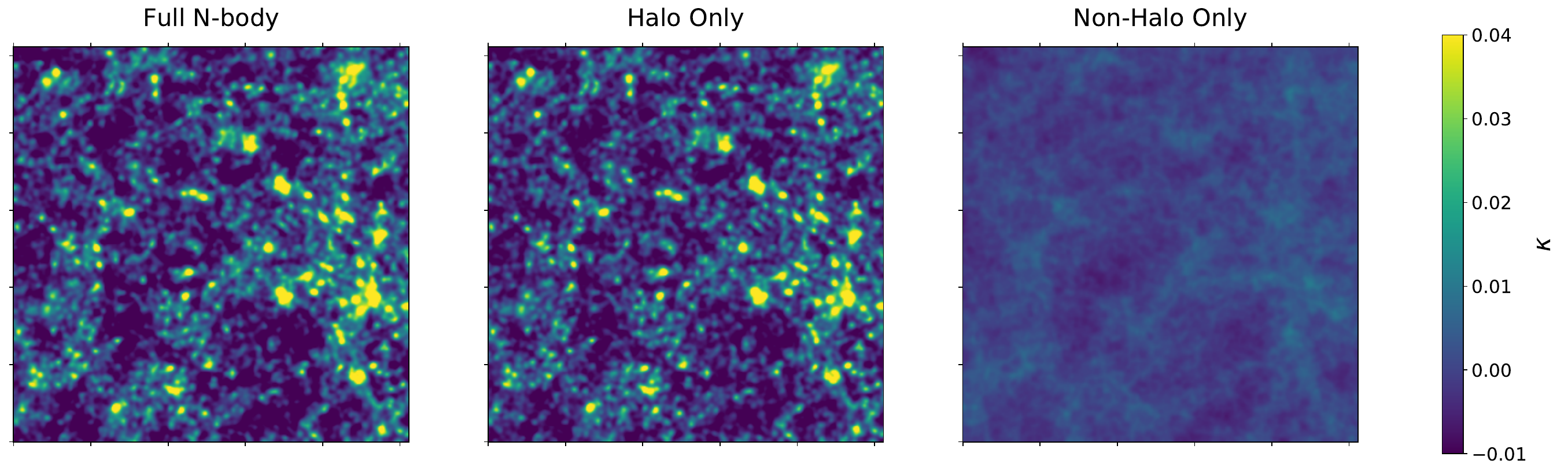}
\caption{\label{fig:maps} Decomposition of a smoothed convergence map in the fiducial cosmology into halo and non-halo contributions. The visual near indistinguishability of the full N-body and halo maps suggest that the highly significant peaks should be well captured by the halo contribution. The non-halo maps still trace somewhat similar structures, arising from overdensities in the outskirsts of halos.}
\end{figure*}

Next, we generate new sets of snapshots, in which we include only particles that belong to halos as identified by \texttt{Rockstar}.  We process these snapshots using \texttt{Lenstools} in the same way as the full N-body runs, described in \S~\ref{methodology}.A:  we create lensing planes, and ray-trace through them to produce ``halo-only'' convergence maps, representing the contributions of DM halos.   Note that this procedure does not conserve mass, as the non-halo material is effectively removed.  Equivalently, this corresponds to the assumption that the non-halo material is smoothly distributed outside halos, and does not contribute to the local convergence in the weak lensing limit~\cite{KaiserSquires1993}.
We also generate analogous ``non-halo-only'' maps, following the same procedure, except we select all particles that are not tagged as halo particles by \texttt{Rockstar}. Fig.~\ref{fig:maps} shows as an illustration the decomposition of a noiseless convergence map into its halo-only and non-halo-only contributions.   The figure shows that the ``halo-only'' maps look very similar to the full N-body maps, and that the non-halo-only maps still trace somewhat similar structures, arising from overdensities in the outskirts of halos.

\subsection{Statistical Inference Methods}

We use Bayes' theorem to infer the values of cosmological parameters from the peak counts measured on our maps. For a model M (defined by our full simulation pipeline), a data vector $\textbf{x}^{\rm obs}$ (binned peak counts) and a set of parameters $\theta$ (${\Omega_{\rm m}, \sigma_{8}}$), the posterior probability distribution $p$ for the parameters is given by
\begin{equation}
    p(\theta|\textbf{x}^{\rm obs},M)=\frac{p(\textbf{x}^{\rm obs}|\theta, M)p(\theta, M)}{p(\textbf{x}^{\rm obs},M)}.
\end{equation}

\begin{table}[t]
\caption{\label{tab:table3} Example mean peak counts (averaged over 500 map realizations) in the fiducial cosmology ($\Omega_{\rm m}=0.26$, $\sigma_{8}=0.8$). S/N bin edges for our 8-bin observable are  $[-\infty,-1,0,1,2,3,3.5,4, +\infty]$, where noise level is defined as the r.m.s. $\sigma_{\kappa}=0.02$ (see \S~\ref{methodology}). 
}
\begin{ruledtabular}
\begin{tabular}{lcc}

Maps & Peak Counts\\
\\
Full N-body & [ 7.28,170.41,707.69 \\
&686.12, 223.70,30.78,13.35,14.36] \\
\\
Halo-Only & [5.00,154.45,725.33, \\ 
& 712.40,214.10,27.29,11.71,12.70]\\
\\
\end{tabular}
\end{ruledtabular}
\end{table}

In our analysis, the observable, $\textbf{x}^{\rm obs}$, is a histogram of peak counts, i.e. the number of peaks on a simulated map at each given S/N. We use 8 S/N bins as listed in Table~\ref{tab:table3}. We chose to use 8-bin histograms to ensure that the observables' covariance is non-singular for all cosmological models and all map sets (full N-body and halo-only). In our analysis, 
all the maps share the same underlying generative model (ray-traced N-body simulation of a Lambda Cold Dark Matter ($\Lambda$CDM) cosmology)
and we therefore drop the model-dependent evidence term, $p(\textbf{x}^{\rm obs}$,M), as it is constant. 
We set the parameter priors, $p(\theta, M)$, to zero outside the parameter space that we are exploring and uniform inside ($\theta \in$ \{$\Omega_{\rm m},\sigma_{8}$\}, where $0.160<\Omega_{\rm m}<0.601$ and $0.150<\sigma_{8}<1.251$). This allows us to also leave out the (constant) prior term in our analysis. Our posterior distribution, therefore, equals the likelihood up to a constant factor:
\begin{equation}
    p(\theta|\textbf{x}^{\rm obs}) \propto  p(\textbf{x}^{\rm obs}|\theta)\propto L(\theta)
\end{equation}

By assuming that our observable $\textbf{x}^{\rm obs}$ follows a multivariate Gaussian distribution~\cite{Gupta+2018}, we can use the Gaussian log-likelihood function.
It has previously been shown that a Gaussian likelihood is a good approximation for weak lensing peaks and does not result in significant changes to credible contours (\cite{Gupta+2018}, ZM16, \cite{Lin2015b}):
\begin{equation}
    \log L (\theta)=\log \left[(2d)^{d}\det(C(\theta))\right]+ \Delta\textbf{x}(\theta)^{T}\hat{C}^{-1}(\theta)\Delta\textbf{x}(\theta)
\end{equation}
where $\Delta\textbf{x}$ is the difference between the mean peak counts for a given cosmology and those measured at the fiducial cosmology ($\Omega_{\rm m}=0.260$, $\sigma_{8}=0.800$) in full N-body maps,
\begin{equation}
    \Delta\textbf{x}=\bar{\textbf{x}}(\theta)-\bar{\textbf{x}}(\theta_{\rm fid})
\end{equation}
and $\hat{C}^{-1}$ is the inverse covariance matrix, re-scaled to the number of realizations ($N=500$) and the number of histogram bins ($d=8$) according to the method in \cite{Hartlap2007} to remove the bias resulting from its estimation,
\begin{gather}
    C(\theta)=\frac{1}{N-1}\sum_{i=1}^{N}(\textbf{x}_{i}(\theta)-\bar{\textbf{x}}(\theta))(\textbf{x}_{i}(\theta)-\bar{\textbf{x}}(\theta))^{T},\\
    \hat{C}(\theta)^{-1}=\frac{N-d-2}{N-1}C(\theta)^{-1}.
\end{gather}
ZM16 have shown that using a cosmology-dependent covariance improves constraints by $14-20$\% and Ref. \cite{Lin2015b} has similarly shown that using a cosmology-independent covariance increases contour area by $22\%$. Since the effect of using cosmology-dependent covariance is comparatively small and the main purpose of this study is to compare constraints from two sets of maps, we fix the covariance to its value at the fiducial cosmology for simplicity, and the log-likelihood in our analysis is therefore simply 
\begin{equation}
    \log L (\theta) \propto  \Delta\textbf{x}(\theta)^{T}\hat{C}^{-1}(\theta_{\rm fid})\Delta\textbf{x}(\theta)
\end{equation}

We have N-body simulation outputs for 97 cosmologies. Therefore, in order to evaluate the likelihood at a point in parameter space for which no N-body simulation is available, we need to interpolate the data vector. After testing several interpolating functions from the Python's SciPy library \cite{2020SciPy}, we chose the 2-D Clough-Tocher cubic interpolator, and applied it to each element of our data vector independently. As an accuracy test, we estimated the peak histogram for the fiducial model, interpolated using the remaining 96 cosmologies.
Residuals between the interpolation and true values are all below $10^{-2}\sigma$ with a median value across eight bins of $\approx 6 \times 10^{-5}\sigma$ where $\sigma$ is the r.m.s. in each bin over the 500 realizations of the fiducial model. The difference is small, so we do not expect interpolation to significantly affect our parameter inference. To build credible intervals, we interpolate over the region that spans $\Omega_{\rm m} \in [0.160, 0.600]$ and $\sigma_{8} \in [0.150, 1.250]$ with 4,000 evenly spaced values for each parameter. We have checked that using a finer interpolation grid does not have a significant effect on the contour areas or shapes. 

\section{Results}

\begin{figure*}[!ht]
\includegraphics[width=\textwidth]{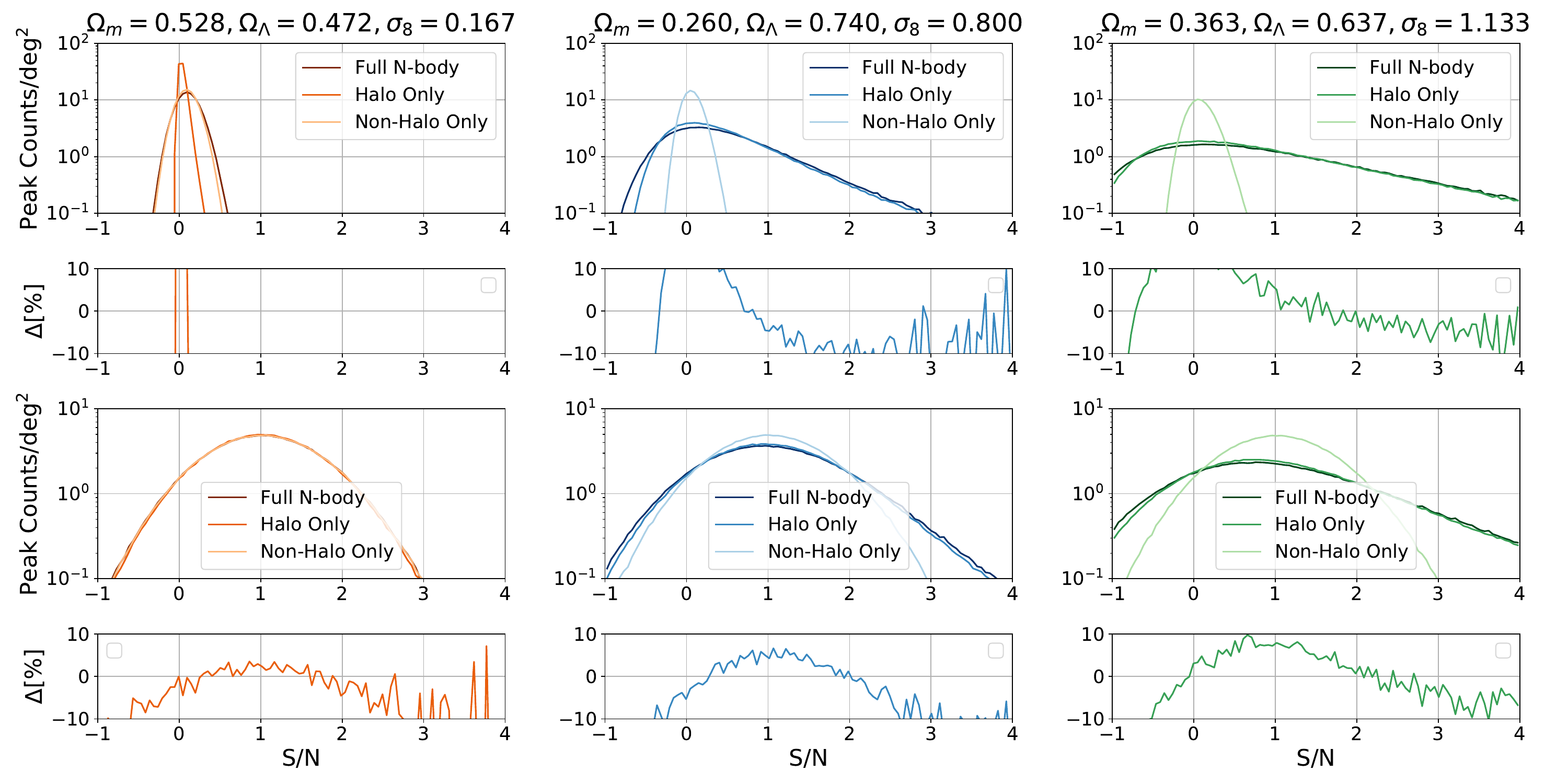}
\caption{\label{fig:peakhist} Mean peak counts in three different cosmologies: the fiducial (middle column), a low-$\Sigma_{8}$ (left column) and a high-$\Sigma_{8}$ (right column) cosmology. The top pair of panels shows mean peak counts averaged over 500 smoothed noiseless WL maps, while the bottom pair of panels shows peak counts after noise is added. 
The histograms are based on 100 bins. The second and fourth row of panels show the fractional differences between full N-body and halo-only maps 
($\Delta=(N_{\rm halo} - N_{\rm full})/N_{\rm full}$).}
\end{figure*}

Fig. \ref{fig:peakhist} shows the mean peaks counts as a function of their significance level for convergence maps ray-traced from full N-body simulations, halo-only and non-halo-only contributions. As an illustration, we plot the peak histograms for three cosmologies: low-$\Sigma_{8}$, fiducial and high-$\Sigma_{8}$ (picked from our available cosmologies as shown in Fig. \ref{fig:grid}). Here $\Sigma_{8}$ is a 
derived parameter that is roughly orthogonal to the well-known $\Omega_m-\sigma_8$ degeneracy direction
($\Sigma_{8}=\sigma_{8}\left(\frac{\Omega_{\rm m}}{0.3}\right)^{0.6}$, as used in ZM16). In Fig. \ref{fig:peakhist} we use for clarity 100 equally-sized bins (except two bins on the edges) in the S/N interval $[-\infty, -1.0,...,4.0,+\infty]$. The top pair of panels shows peak counts and residuals from averaging 500 smoothed WL maps, while the bottom pair of panels correspond to averaged peak counts from 500 WL maps that were both smoothed and combined with noise maps as described in \S~\ref{methodology}. 

From the top pair of panels, it can be seen that in the fiducial, and especially in the high-$\Sigma_8$ model, halo-only maps are able to reproduce positive peak counts reasonably well, with $< 10\%$ errors.  On the other hand, they more significantly over-predict peaks near S/N $\approx$ 0, and under-predict negative peaks.   Halo-only predictions are much more discrepant in the low-$\Sigma_8$ model.

Once noise is applied, peak counts in the full N-body and halo-only maps become somewhat closer, but their differences still exceed several percent across most bins. The median differences are $\approx 4\%$, $5\%$, and $5\%$, and the maximum differences are $\approx 56\%$, $31\%$, and $32\%$ for the three cosmologies shown.

\begin{figure}[ht]
\includegraphics[width=\linewidth]{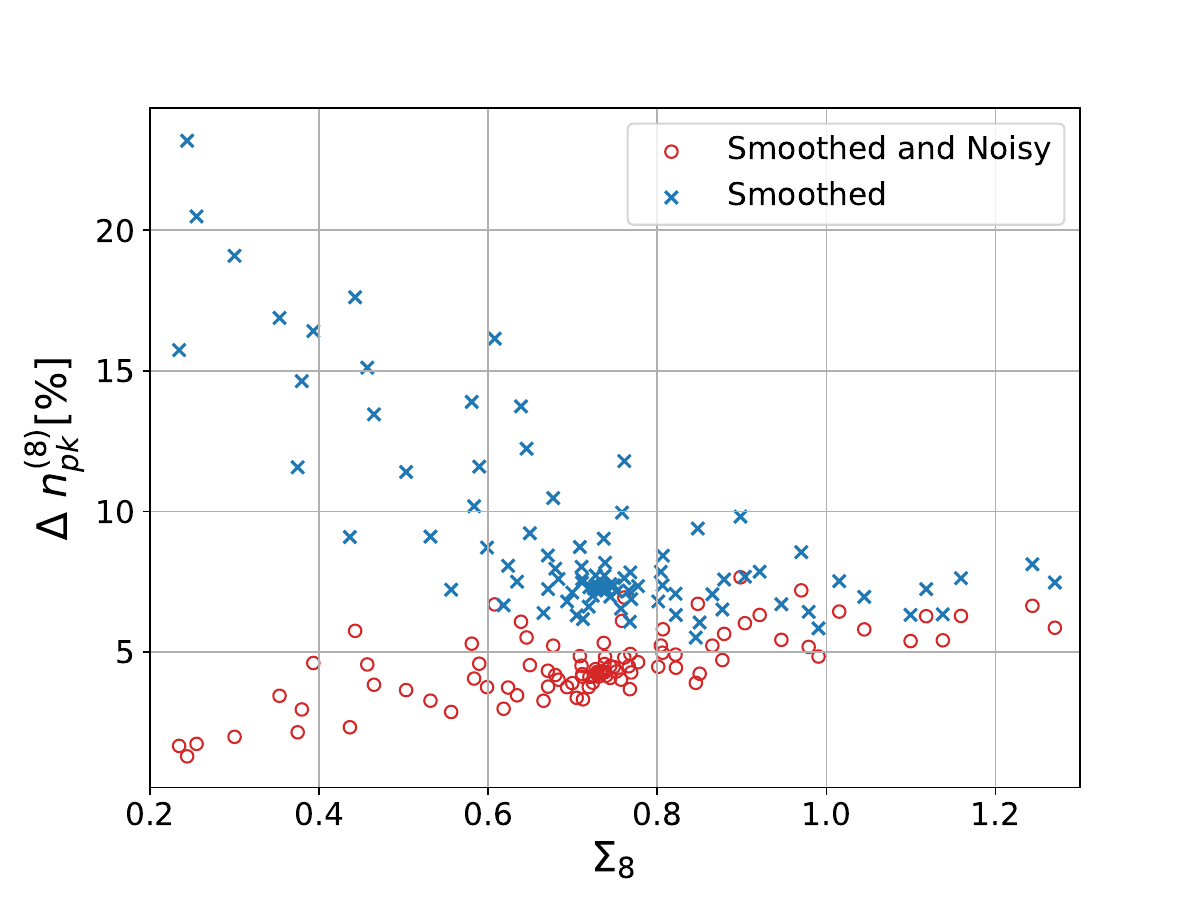}
\caption{\label{fig:histdiff} Average percentage difference in peak counts as a function of the best-constrained parameter combination $\Sigma_{8}\equiv\sigma_{8}\left(\frac{\Omega_{\rm m}}{0.3}\right)^{0.6}$. Here $\Delta n^{(8)}_{\rm pk}=\sum_{i=1}^{n=8} |N_{{\rm halo},i} -N_{{\rm full},i}|/\sum_{i=1}^{n=8}N_{{\rm full},i}.$}
\end{figure}

In Fig.~\ref{fig:histdiff} we also compare the 8-bin histograms ($n^{(8)}_{\rm pk}$) from smoothed and noisy maps by plotting the average difference in total peak counts across bins as a function of $\Sigma_8$. As might be intuitively expected, and also seen in Fig.~\ref{fig:peakhist}, in the absence of noise, halos contribute less to peaks in cosmologies with lower $\Sigma_8$, which have less evolved nonlinear structures, leading to larger discrepancies between the halo-only and full maps. However,
as was shown in ZM16, once noise is added, it dominates in cosmologies with low $\Sigma_{8}$, diluting these otherwise large differences.

To study how cosmological parameter estimates are impacted by the halo-only predictions, we compute the log-likelihoods as described in \S~\ref{methodology}, setting our mock data vector and covariance matrix to the ones derived from the full N-body maps in our fiducial cosmology. We then compute $68.3\%$ and $95.4\%$ confidence contours using peak-count predictions from either the full N-body or the halo-only maps. Since our parameter space is only 2-D, it is computationally feasible to evaluate the likelihoods directly on a grid, instead of sampling them using Markov chain Monte Carlo methods. 

As our mock data vector is taken from the full N-body maps, we expect to see a bias in the contours when halo-only maps are used for parameter inference. For our original-sized maps ($3.5\times3.5\,\text{deg}^2 \approx 12\,\text{deg}^2$), we find that the unbiased and biased confidence intervals overlap but have different tilts and are slightly offset from each other. To scale our results to larger survey areas, we assume that our maps are representative of an average region within a survey, so that our log-likelihoods increase proportionally to the survey area. As we scale our log-likelihoods for larger surveys, the bias in the contours becomes more significant.

\begin{figure*}[!ht]
\begin{centering}
\includegraphics[width=\textwidth]{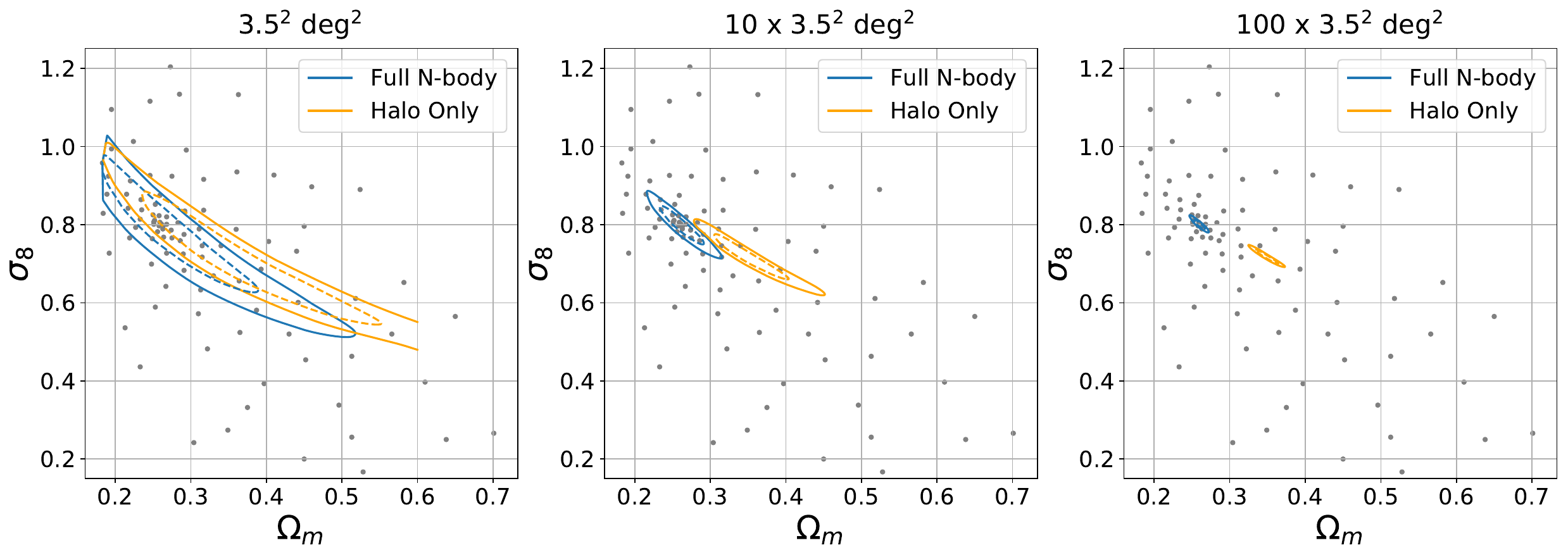}
\caption{\label{fig:contourspeaks} Comparisons of the $68.3\%$ and $95.4\%$ credible contours obtained from peak counts, when maps from the full N-body simulations (blue) or from those that include halo particles only (orange) are used to fit peak counts in the mock data from the full fiducial N-body simulations. The observable is the 8-bin peak count histogram with S/N edges 
$[-\infty,-1,0,1,2,3,3.5,4, +\infty]$ and the fiducial cosmology corresponds to $\Omega_{\rm m}=0.26$ and $\sigma_{8}=0.8$. Log-likelihoods were scaled from the original maps (size $\approx$ 12 deg$^{2}$; left panel) to $\approx$ 123 deg$^{2}$ (middle) and $\approx$ 1,225 deg$^{2}$ (right). Using the halo-only maps for inference results in significantly biased cosmological constraints for surveys $\gtrapprox$ 100 deg$^{2}$.}
\end{centering}
\end {figure*}

Fig. \ref{fig:contourspeaks} shows how the contours for maps with areas 10 or 100 times larger than our original map size ($\approx$ 123 deg$^{2}$ and $\approx$ 1,225 deg$^{2}$), no longer overlap. Current and future surveys are much larger than the size above which the confidence intervals no longer overlap (i.e. DES-wide\footnote{https://www.darkenergysurvey.org/}: 5000 deg$^{2}$,  Euclid\footnote{https://sci.esa.int/web/euclid}: 15,000 deg$^{2}$, LSST \footnote{https://www.lsst.org/}: 18,000 deg$^{2}$). Thus, we can expect to obtain cosmological constraints from halo-only WL maps that are strongly biased. In Table \ref{tab:areacentrs} we list percent differences in areas and centroid positions of credible contours as compared to the full N-body maps. We calculate the centroid positions as follows,
\begin{equation}
    \theta^{\rm centroid}_{i}=\frac{\int_{\rm CR}L(\theta_{i}, \theta_{j})\theta_{i}d\theta_{i}d\theta_{j}}{\int_{\rm CR}L(\theta_{i}, \theta_{j})d\theta_{i}d\theta_{j}},
\end{equation}
where $\theta_{i}$ refers to the axis along which the centroid position is computed, $\theta_{j}$ is the orthogonal axis in our parameter space, and CR refers to the contour region over which the integration is performed. For our original 12 deg$^{2}$ maps, the $95.4\%$ contour areas of halo-only maps are $\approx 11\%$ larger and have centroid positions that are shifted $+31\%$ and $-8\%$ along $\Omega_{\rm m}$ and $\sigma_{8}$, respectively. 

To quantify the shift of the contours in the direction along and perpendicular to the best-constrained parameter combination $\Sigma_{8}=\left(\frac{\Omega_{\rm m}}{0.3}\right)^{\alpha}$, as well as their  tilt, we follow the method outlined in Ref.~\cite{Petri2015} to find the best-fit $\alpha$. Using the entire grid of interpolated values, we find $\alpha\approx0.58$ and $\alpha\approx0.55$ for full N-body maps and halo-only maps, respectively. To directly assess the bias, we use $\alpha=0.56$ and the most likely $\Omega_{\rm m}$ and $\sigma_{8}$ values in the two sets of maps.  

We find the bias to be $\approx 42\%$ along the direction of degeneracy and $\approx 6\%$ perpendicular to it. As also seen visually in Fig.~\ref{fig:contourspeaks}, most of the bias is along the direction of degeneracy. We also find that using the contours from the halo-only maps, the fiducial model lies at the contour enclosing $\approx 69.6\%$ of the total likelihood in the $\approx 12$ deg$^{2}$ case (i.e., no strong bias).  However, the fiducial model is strongly biased, and beyond the 99\% contours in the $\approx$ 123 deg$^{2}$ and $\approx$ 1,225 deg$^{2}$ maps. Marginalizing over $\sigma_{8}$, we find that the fiducial $\Omega_{\rm m}$ value lies at $\approx 1 \sigma, 3 \sigma,$ and $9 \sigma$ for the three different sized surveys. Similarly, the fiducial $\sigma_{8}$ is at $\approx 1 \sigma, 2 \sigma,$ and $7 \sigma$.

\begin{table}[!hb]
\caption{\label{tab:areacentrs} Percentage differences in $(\Omega_{\rm m},\sigma_8$) contour areas when halo-only maps vs. full N-body maps are used for parameter inference. The first column is for our fiducial maps and the next two columns show results for 10 and 100 times larger areas.}
\begin{ruledtabular}
\begin{tabular}{lccc}
& $3.5^{2}$ deg$^{2}$ & 10$\times$($3.5^{2}$ deg$^{2}$) & 100$\times$($3.5^{2}$ deg$^{2}$)\\
\colrule

$68.3\%$ & +25 & +46 & +78 \\
$95.4\%$ & +11 & +41 & +58 \\

\end{tabular}
\end{ruledtabular}
\end{table}

\begin{figure*}[!ht]
\includegraphics[width=\textwidth]{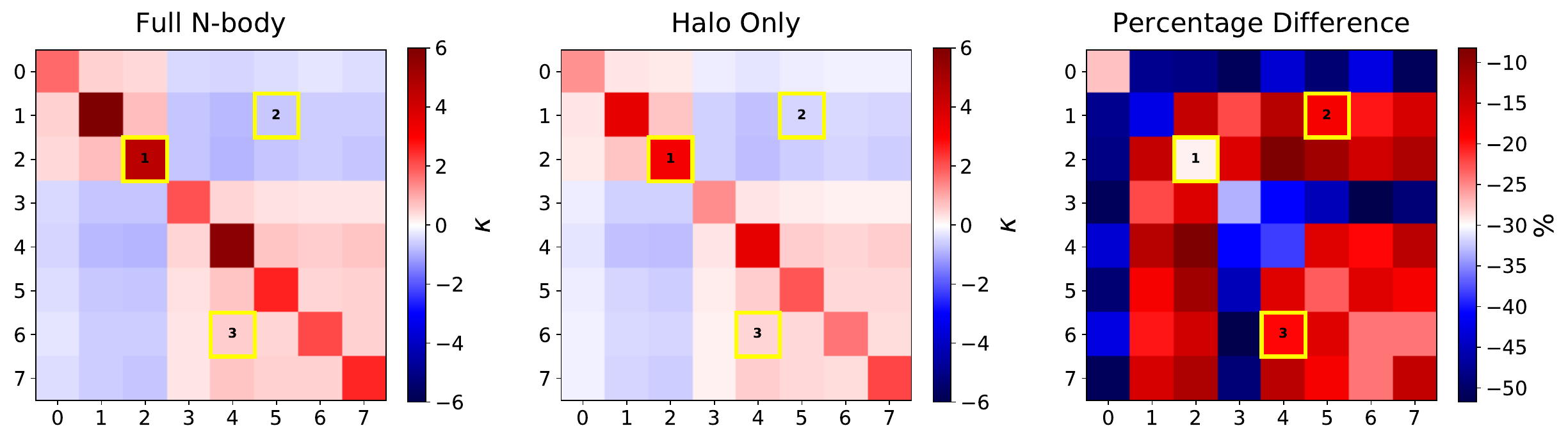}
\caption{\label{fig:corrfid} Correlation matrices for peak counts in the fiducial cosmology ($\Omega_{\rm m}$=0.26, $\sigma_{8}$=0.8). Diagonal elements were replaced by the peak count variance divided by its mean ($\sigma^{2}_{ii}/\bar{x}_{ii}$). Marked matrix elements are plotted for each cosmology in Fig. \ref{fig:corrfid}.}
\end{figure*}

\begin{figure*}[!ht]
\includegraphics[width=\textwidth]{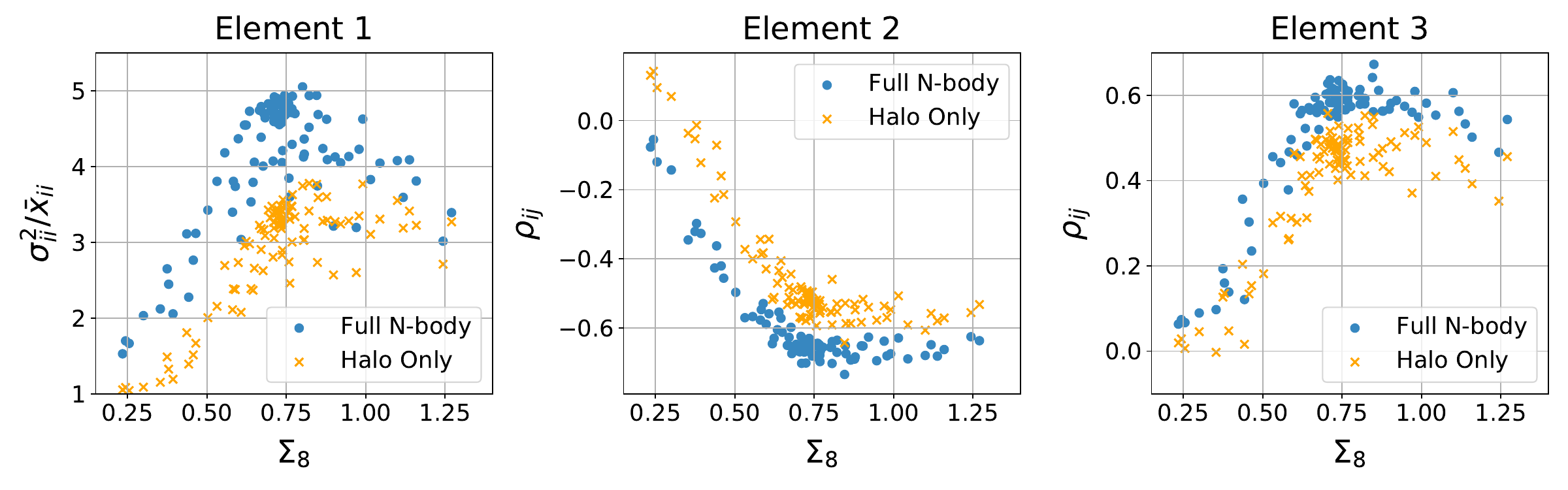}
\caption{\label{fig:corrall} Selected correlation matrix elements are shown as a function of the best-constrained parameter-combination $\Sigma_8$. Halo-only maps do not account for all the variance (the median difference for element 1 is $\approx 30\%$) or correlations (median differences for elements 2 and 3 are $\approx 20\%$); they also fail to fully capture their cosmology-dependence.}
\end{figure*}

To further quantify the differences between peaks in the halo-only and full N-body WL maps, we compare their correlation matrices. Fig.~\ref{fig:corrfid} shows these matrices for the fiducial cosmology for full N-body and halo-only maps, with diagonal terms replaced by the variance terms normalized by the mean peak counts ($\sigma^{2}_{ii}/\bar{x}_{ii}$). Both full N-body and halo-only correlation matrices show that high peaks and low peaks are correlated amongst themselves but anti-correlated with one another as was shown in ZM16. Full N-body has overall higher absolute values along all matrix elements. However, unlike in the previous comparisons with CAMELUS in ZM16, the discrepancies here are significantly smaller. The median difference between the absolute values of the corresponding elements in the two matrices is 22\%, with the halo-only maps underpredicting the (co)variances. In Fig. \ref{fig:corrall} we plot three elements of the matrix for all cosmologies: one element each from the diagonal, negative-correlation and positive-correlation regions. Both full N-body and halo-only maps show a strong dependence on cosmology, in contrast with the cosmology-insensitivity of CAMELUS (co)variances shown in ZM16. ZM16 suggested that CAMELUS can be improved by taking into account halo clustering instead of placing halos randomly, which would increase covariances. Here, we see that covariances for halo-only maps are indeed much closer to full N-body maps. However, the halo-only maps still do not account for the total variance based on element 1 in Fig. \ref{fig:corrall} where the median difference is $\approx$ 30\% and maximum difference is 47\% over $\Sigma_{8}$. Halo-only maps account more for the positive and negative correlations as seen for elements 2 and 3, for which median differences over all cosmologies are 20\%. 

\section{Discussion}\label{discussion}

Our results on peak counts based on halo-only maps, and the corresponding confidence contours, have shown that halos alone cannot account for lensing peaks at the accuracy required for on-going and future surveys. Since halos are perfectly modeled in this analysis, this result is robust, and implies that non-halo material is important and must be included in WL peak predictions.

We therefore begin this section by investigating whether material outside but associated with halos may explain the remaining differences in peak counts (A).   In the rest of this section, we then discuss several other issues, including the nature of negative peaks (B), the separate impact of halo clustering and halo shapes on the peak counts (C), the related statistic of lensing minima (D) and caveats about our methods and conclusions (E).

\subsection{Effects from Non-Halo Contributions}

In this section, we assess the impact on the peak counts of including mass from outside halos.

\subsubsection{Marginal Halos}

\begin{figure*}[ht]
\includegraphics[width=\textwidth]{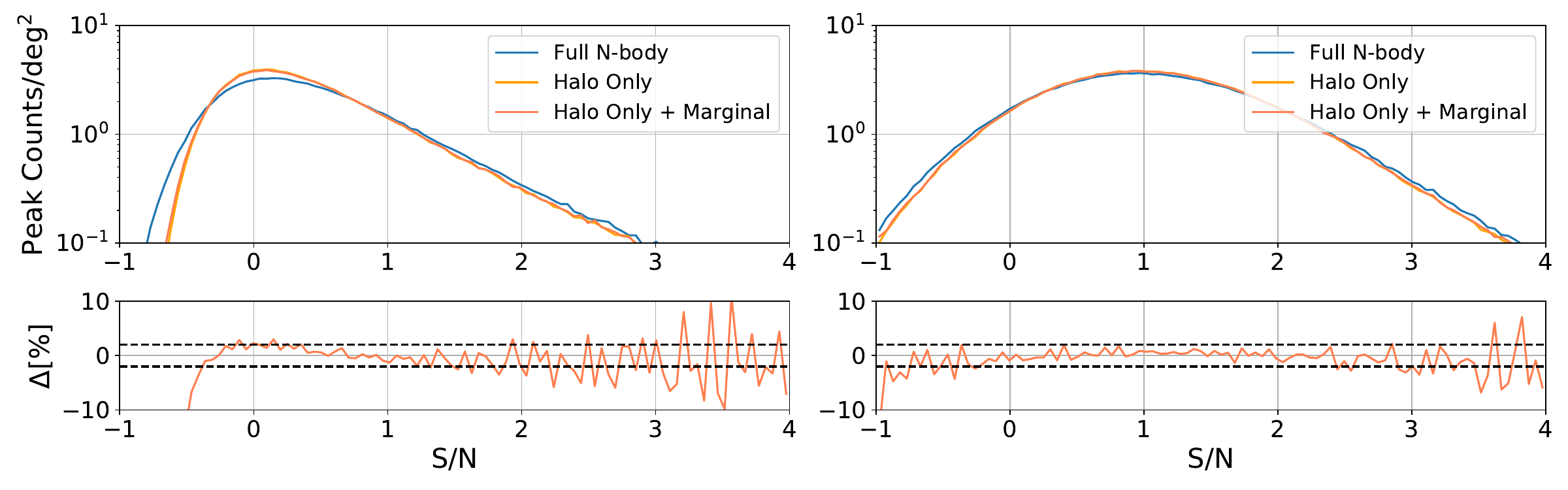}
\caption{\label{fig:marginnomargin} Peak histograms for full N-body, halo-only and halos + marginal halos maps. The left pair of panels corresponds to peak histograms from smoothed noiseless maps, while the right from smoothed maps combined with noise. Bottom panels represent percentage difference between halo-only and halo + marginal halos maps ($\Delta=(N_{\rm halo}-N_{\rm halo+marginal})/N_{\rm halo}$), where dashed lines mark $2 \%$ difference. The additional contribution from marginal halos does not significantly affect halo-only peak histograms. 
}
\end{figure*}

To construct our halo-only maps, we used particles that belong to halos in the main halo catalogs produced by \texttt{Rockstar}. However, \texttt{Rockstar} additionally identifies particles that belong to marginal halos that have not reached a high enough S/N ratio to be included in the main halo catalog. These marginal halos include small halos at the mass resolution limit, as well as massive nonlinear structures that do not reach a sufficiently high overdensity. To check the effect of incorporating these marginal halos, we compare peak histograms in the fiducial cosmology with and without these objects included. Although marginal halos add $\approx 8-13 \%$ more particles to our halo-only snapshots, the effect on either the 100-bin or the 8-bin histograms is small. The median and maximum absolute percentage differences between halo-only and (halos + marginal halos) maps in the 100-bin vs. the 8-bin histograms from noisy maps are $\approx 1\%$ and $ 13\%, \mathrm{vs. }1\%$ and $3\%$, respectively.
The differences for 100-bin histograms are shown in Fig.~\ref{fig:marginnomargin}. We also computed confidence contours using maps that include marginal halos. We found no visible differences, although the $95.4\%$ contour area is increased by $\approx 1\%$ for $3.5 \times 3.5$ deg$^{2}$ maps, and by $\approx 7\%$ for $100 \times (3.5 \times 3.5$ deg$^{2}$) maps.

\subsubsection{Near-Halo Contributions}

Another hypothesis is that mass residing near the outskirts of halos may account for most of the remaining discrepancies.
To investigate the contribution of this extra material to the peak counts, we look at the histograms computed from simulation snapshots that include all halo particles, plus all particles within different numbers \textit{n} of virial radii of their closest halo. 

\begin{figure*}[!t]
\includegraphics[width=\textwidth]{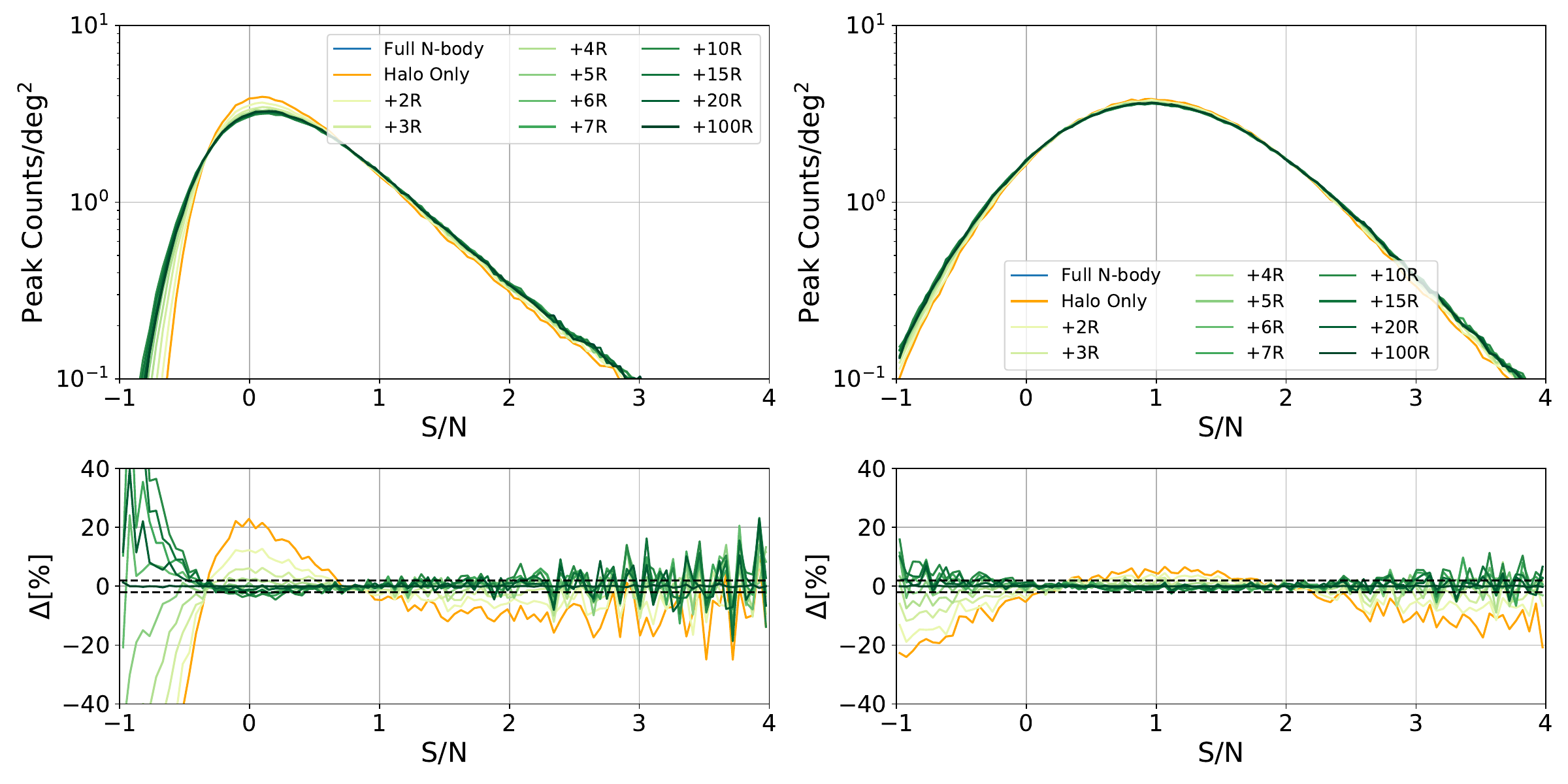}
\caption{\label{fig:allR_diff} 100-bin peak count histograms from full N-body, halo-only and (halos + particles within $nR_{\rm vir}$ of halos) maps. As more material farther away from halos is included, negative peak counts increase but medium- and high-S/N peak counts decrease. The residuals converge to zero as $n$ is increased, but the convergence is very slow. The horizontal dashed lines mark $\pm 2\%$ differences for reference. \footnote{Small fluctuations in residuals from $100R_{\rm vir}$ are due to numerical noise (truncated values of particle positions in newly generated snapshots). This also applies to Fig.~\ref{fig:mass_hists}.}
}
\end{figure*}

To do this, we place a sphere centered on each halo, with a radius corresponding to \textit{n} times the virial radius $R_{\rm vir}$ of that halo as determined by \texttt{Rockstar}, and catalog all particles residing in these spheres.  Fig.~\ref{fig:allR_diff} shows 100-bin histograms for the full N-body, original halo-only and (halos+outskirts) maps; the latter including the outskirst to $nR_{\rm vir}$ with $n=[2,3,4,5,6,7,10,15,20,100]$.  As we include material farther away from the halos, the number of negative peaks increases and the negative part of the histogram matches that of the full N-body maps better. However, as we increase $n$, the medium-height peak counts decrease and the negative part of the histogram is still not reproduced exactly until we include very high $n$. In order to be within $\approx 2\%$ residuals in our 8-bin observable, we find that we need to include particles within at least $\approx 15R_{\rm vir}$. For reference, particles within one virial radius of their nearest host halo occupy $\approx 0.1\%$ of our simulation volume at $z=0$.

\begin{figure}[!h]
\includegraphics[width=\linewidth]{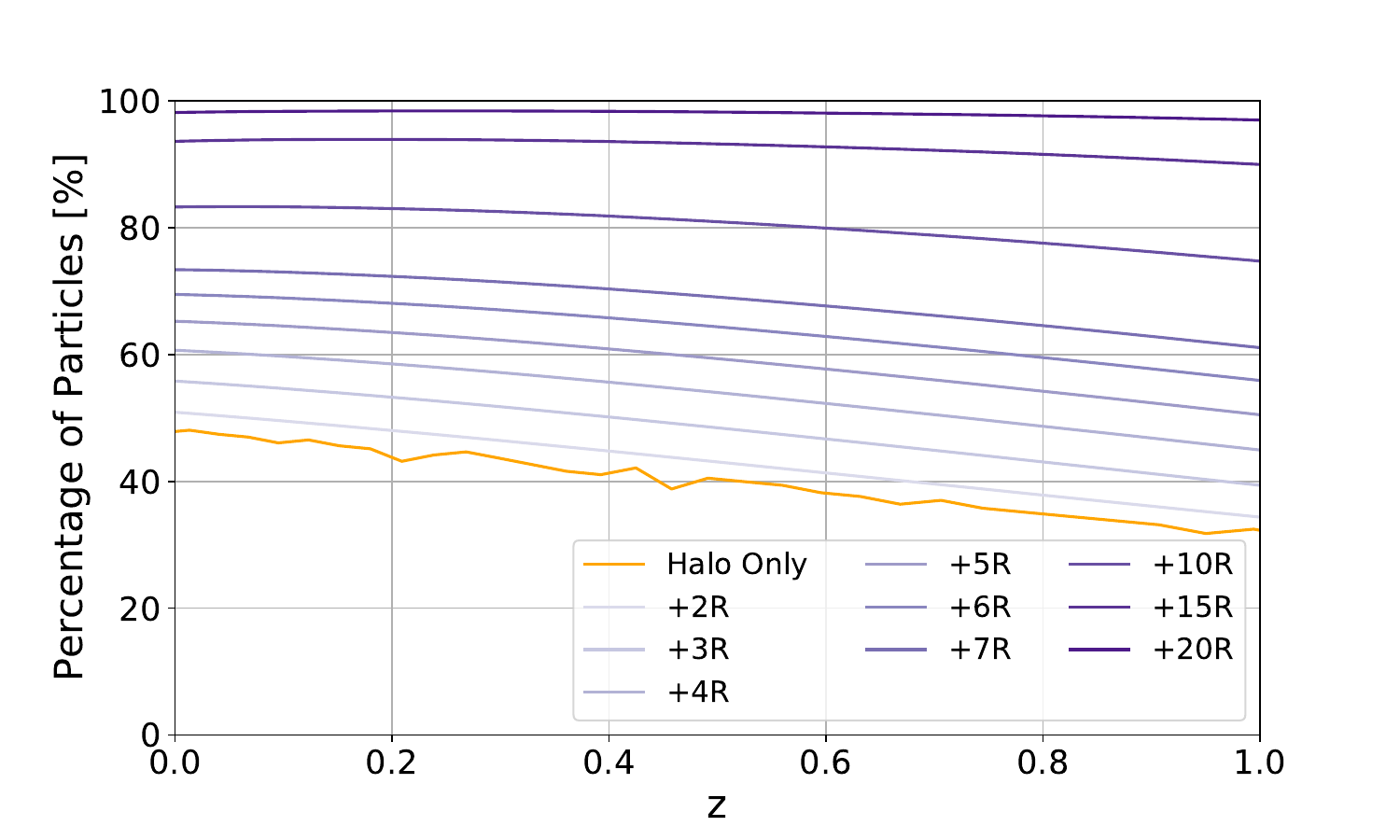}
\caption{\label{fig:mass_z} Percentage of particles (or equivalently, the total mass fraction) as a function of redshift within different distances from halos (i.e. different $n$ in $nR_{\rm vir}$). Simulation snapshots at larger redshifts generally include less mass as nonlinear structure formation is at a less advanced stage.
The difference between the closest and farthest snapshots decreases with larger $n$.}
\end{figure}

\begin{figure}[!h]
\includegraphics[width=\linewidth]{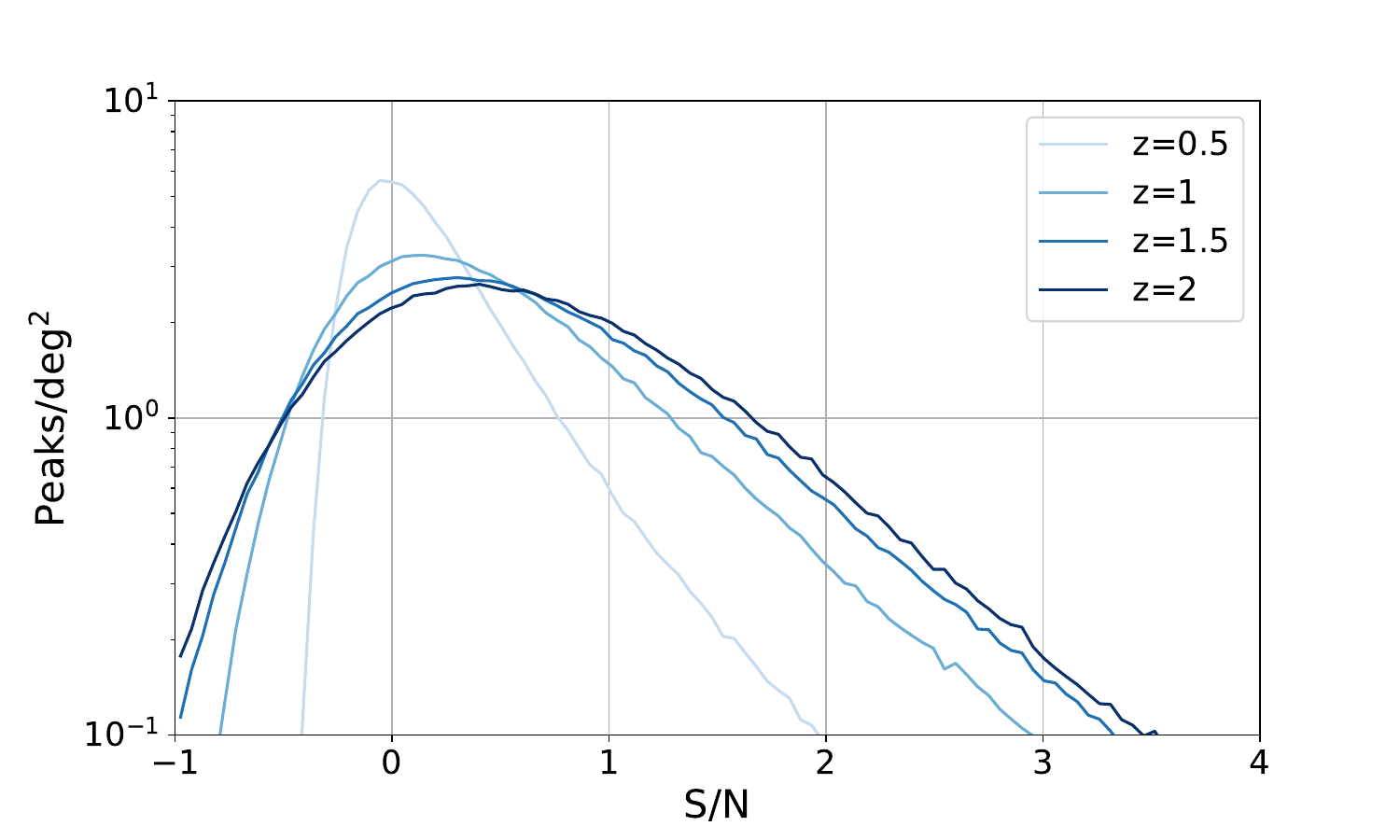}
\caption{\label{fig:full_z} Smoothed 100-bin peak count histograms from full N-body simulations with source galaxies placed at $z=0.5, 1, 1.5$ and 2. Since the r.m.s. convergence increases with higher redshift, we use r.m.s = 0.017, 0.020, 0.023, 0.026 as the noise N in the S/N at the four redshifts, respectively. The distribution of negative, medium- and high-significance peaks is redshift-dependent, with more negative and high-S/N peaks when material from higher redshift is included.}
\end{figure}

Fig. \ref{fig:allR_diff} shows that including material far away from halos is important for O(1\%) accurate peak counts. The lack of a rapid convergence of the residuals to zero as $n$ is increased can partly be explained by the fact that a different fraction of particles are included at different redshifts. Halos are rarer and contain a smaller fraction of the total mass at higher redshift and therefore the mass fraction in halo particles + $nR_{\rm vir}$ vary as a function of redshift. We show these fractions in Fig.~\ref{fig:mass_z}. This means that non-halo material becomes even more important at higher redshift and needs to be incorporated into peak count models for deeper surveys. This can be clearly seen, alternatively, by examining peaks on maps raytraced to source galaxies placed at increasing redshifts. Fig.~\ref{fig:full_z} shows 100-bin histograms from smoothed maps for full N-body simulations with source galaxies at $z =0.5, 1, 1.5$ and $2$. The figure shows the effect of including material at high vs. low redshift and how including only material at lower redshifts results in the tails of the histogram removed. When we include halo-only and $nR_{\rm vir}$ particles, we are including fewer particles at high-redshift, yielding histograms which more closely resemble those from raytracing to lower redshifts. 

\begin{figure*}[ht]
\includegraphics[width=\textwidth]{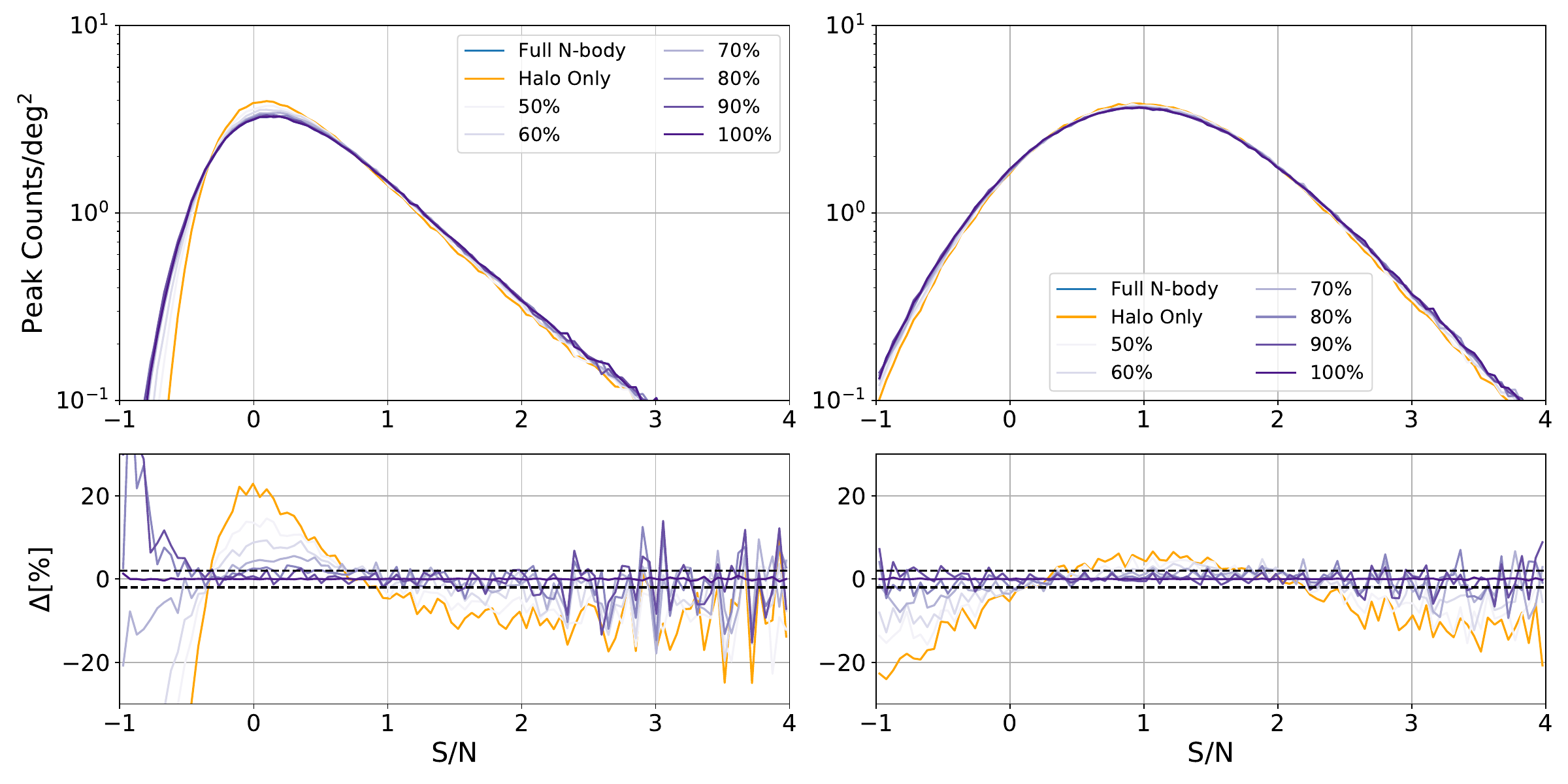}
\caption{\label{fig:mass_hists} 100-bin peak count histograms in full N-body, halo-only and (halos + particles within $nR_{\rm vir}$ of halos) maps, with $n$ adjusted to produce a fixed mass fraction, independent of redshift. As we expect, the residuals converge to zero across all bins as we increase the included mass. The horizontal dashed lines in the lower panels mark $\pm 2\%$ differences for reference.}
\end{figure*}

To remove this difference in the total number of particles across different simulation snapshots, we also look at convergence maps ray-traced from snapshots which have the same fraction of mass included at each redshift. That is, we create new snapshots which include all halo particles plus all particles within some $n$ times $R_{\rm vir}$ of halos, with $n$ adjusted (increased) with redshift to yield a chosen redshift-independent mass fraction. At $z=0$, halo-only maps include $\approx 48\%$ of mass, so in Fig.~\ref{fig:mass_hists}, we look at maps containing 50\%, 60\%, 70\%, 80\%, 90\% and 100\% of the total mass. This figure shows that the residuals converge to zero as more mass is included. However, even when high mass fractions are included, the histograms differ for low-S/N and negative peaks. At least $\approx90\%$ of mass across all redshifts needs to be included to reach a percent-level difference across all bins in our 8-bin observable, using noisy maps. This suggests the influence of voids (and filaments) on peak counts, as those regions fill most of the volume far away from halos. Ref. \cite{Voivodic2020}, for example, has put forward a new model of the large scale structure, which incorporates both voids and dust in addition to halos in the standard halo model formalism. Although this model has only been applied to two-point statistics so far, its effectiveness for predicting WL peaks could be further tested in future work.

\subsection{Inability of Halos to Account for Negative Peaks}

\begin{figure*}[!ht]
\includegraphics[width=\textwidth]{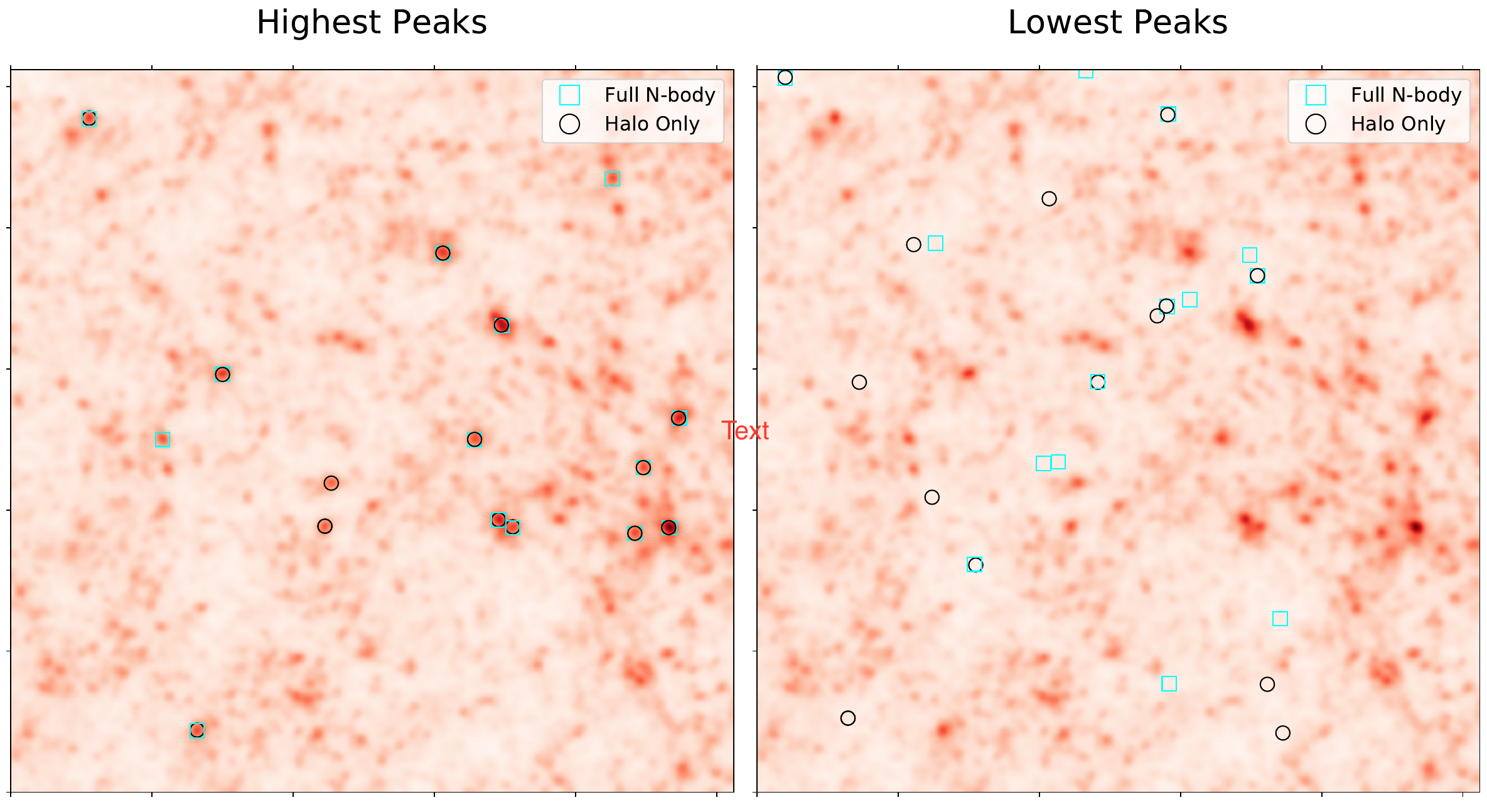}
\caption{\label{fig:posnegpeaks} Example of a single realization of a full N-body convergence map with the 15 highest (left) and 15 lowest (right) peaks marked by squares with 20 pixel ($\approx$ 2 arcmin) sides. Note that the lowest peaks still represent local maxima, but have negative convergence values. 
The black circles mark the positions of the 15 highest and lowest peaks on the corresponding halo-only map.
While positive peaks mostly match in locations between the full N-body and halo-only maps, most negative peaks correspond to different regions. The superposition of halo-only and non-halo material is needed to explain negative peaks.}
\end{figure*}

Fig. \ref{fig:maps} shows visually the differences between convergence maps ray-traced from a full N-body simulation, halo-only and non-halo-only particles for our fiducial cosmology. The maps are consistent with peak histograms we have shown in Fig. \ref{fig:peakhist}, which show that positive peaks are explained well by halo-only maps but negative peaks are not. We inspect individual positive and negative peaks in our maps in Fig. \ref{fig:posnegpeaks}, where we plot the 15 highest and lowest S/N peaks in our maps in order to learn whether they correspond to the same peaks in full N-body and halo-only maps. 
The squares in the images represent roughly 2$\times$2 arcmin areas around the individual peaks, to account for possible shifts in their sky positions due to different levels of lensing deflection near their locations.

Although this visual comparison does not allow us to rigorously compare individual peaks, it demonstrates that most negative peaks in the two sets of maps do not correspond to the same regions in the maps, while the highest (positive) peaks mostly match. Both this and our peak histograms show that the negative peaks, which correspond to under-densities in the (projected) cosmic web, cannot be explained by halo or non-halo material separately but are produced by their superposition along the photon trajectory.

\subsection{Effects of Halo Clustering and Halo Morphology on Peak Counts}

The halo contribution to peak counts shown in this paper results from realistic halos that trace the cosmic web evolved using an N-body code. Alternative halo-based models with lower computational cost rely on a series of approximations. For example, CAMELUS~\cite{Lin2015a, Lin2015b} considers random halo locations, models them with a NFW profiles and assigns them concentrations based only on their mass and redshift.  It is interesting to assess the impact of each of these approximations.  To do so, we generate additional sets of synthetic convergence maps that bridge the gap between our N-body-based, halo-only maps and those generated with the fast halo model CAMELUS. We start by taking our halo-only snapshots and applying the approximations in CAMELUS one at a time.

First, to examine the effect of halo-clustering on peak counts, we generate new maps where each halo position is shuffled randomly. To do this, we group all the particles that share a host halo and shift their position randomly within the 240 Mpc/$h$ simulation volume, preserving the original periodic boundary conditions. 

\begin{figure*}[!ht]
\begin{centering}
\includegraphics[width=\textwidth]{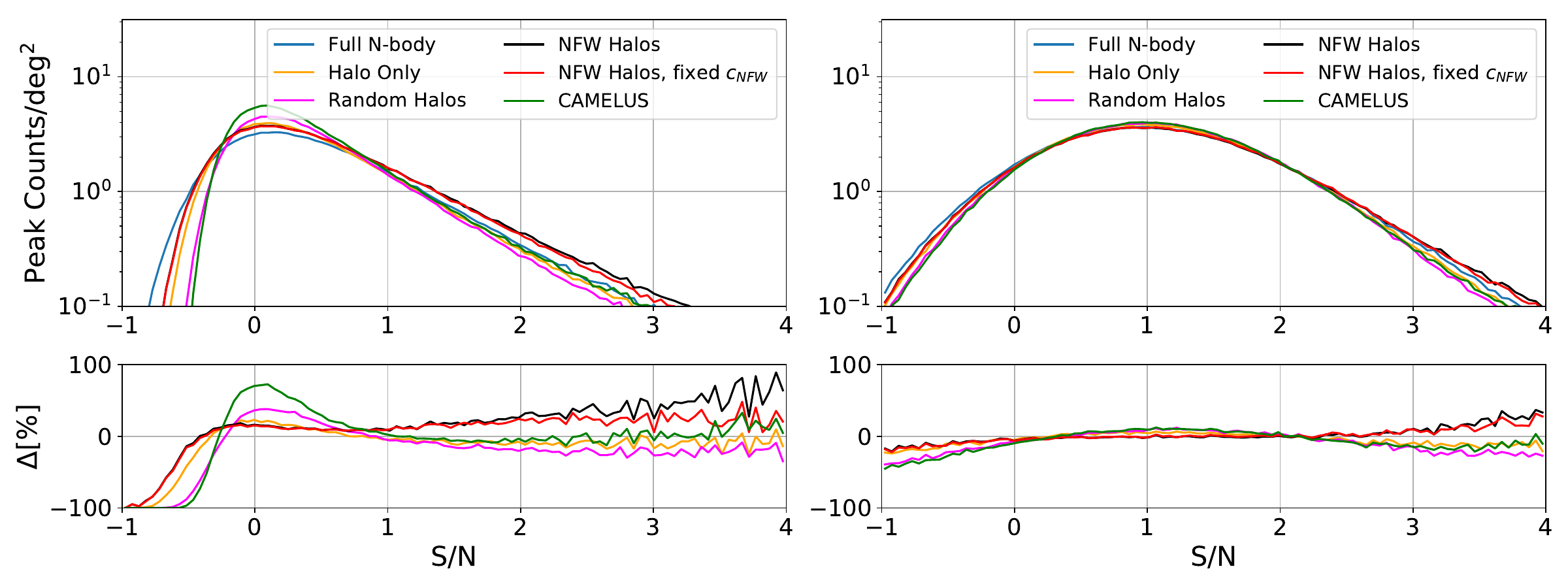}
\caption{\label{fig:disentangled} 
The figure explores the impact of various halo properties on the peak counts.  We show our full N-body simulations and our halo-only predictions (as before), as well as peak counts in the CAMELUS model, which is based on randomly placed NFW halos, with concentration parameters determined by $M_{\rm halo}$ and $z$.   The three remaining curves show peak counts obtained in our N-body simulations, except (i) the halo locations are randomized, to assess the impact of their spatial clustering, (ii) halo profiles are modified to their best-fit NFW shapes, or (iii) halo profiles are modified to NFW shapes with concentration parameters fixed based on their mass and redshift, as in CAMELUS. The left (right) pair of panels shows peak counts from noiseless (noisy) maps. The bottom panel of each pair shows fractional differences relative to the full N-body case.
Halo clustering shifts negative peaks the most, while concentration parameter has a more significant effect on high-S/N peaks.  Note that the two effects partially cancel for S/N$>$1 peaks, making CAMELUS fit these peaks coincidentally quite accurately.}
\end{centering}
\end {figure*}

Next, to assess the effect of using spherical halos, we use NFW density profiles with concentration parameters given by 
\begin{equation}
  c=\frac{10}{1+z}\left(\frac{M_{\rm halo}}{10^{12}\,h^{-1}\mathrm{\rm M}_\odot}\right)^{-0.13},
\end{equation}
in place of halo particles, for all halos with virial radii $R_\mathrm{\rm vir} > 120\,\mathrm{kpc}$, in our potential planes for ray-tracing calculations. The latter choice is due to finite resolution on our potential planes. In this step, we keep the positions of all halos based on the N-body simulations, thus preserving the realistic halo clustering in our potential planes. Lastly, to see the effects of the concentration parameter alone, we repeat the step above except this time the concentration parameter of each NFW profile is set to the actual best-fit value for that halo in the N-body simulation.  We illustrate the effects of these three halo model ingredients, together with peak counts obtained from the CAMELUS model, in Fig.~\ref{fig:disentangled}. 

Looking at the negative part of the histogram, we see that the largest effect on peak counts comes from halo clustering, while the concentration parameter has a smaller effect. This is consistent with ZM16, who found low covariances for CAMELUS and hypothesized that incorporating clustering as a way to improve CAMELUS. 

For positive peaks, we see that randomizing halo positions decreases the peak counts, especially for high-S/N peaks, while having NFW profiles visibly increases positive peak counts.   This increase can be partly explained by the fact that \texttt{Rockstar} does not use the virial radius as a definition for bound halo particles. As such, particles residing outside virial radius can also belong to halos. This means that when we replace halo particles with NFW profiles, we somewhat steepen the density profile, because we ``compressed'' all of the mass inside the virial radius.

An interesting result, revealed in Fig.~\ref{fig:disentangled}, is that for positive peaks, these two effects -- increase due to adopting NFW profiles, and decrease due to ignoring clustering -- tend to cancel.  As a result, CAMELUS predicts, arguably by coincidence, a surprisingly accurate fit for the number of S/N$>$1 peaks, at least in the fiducial cosmology.

\subsection{Weak Lensing Minima}

\begin{figure*}[!ht]
\includegraphics[width=\textwidth]{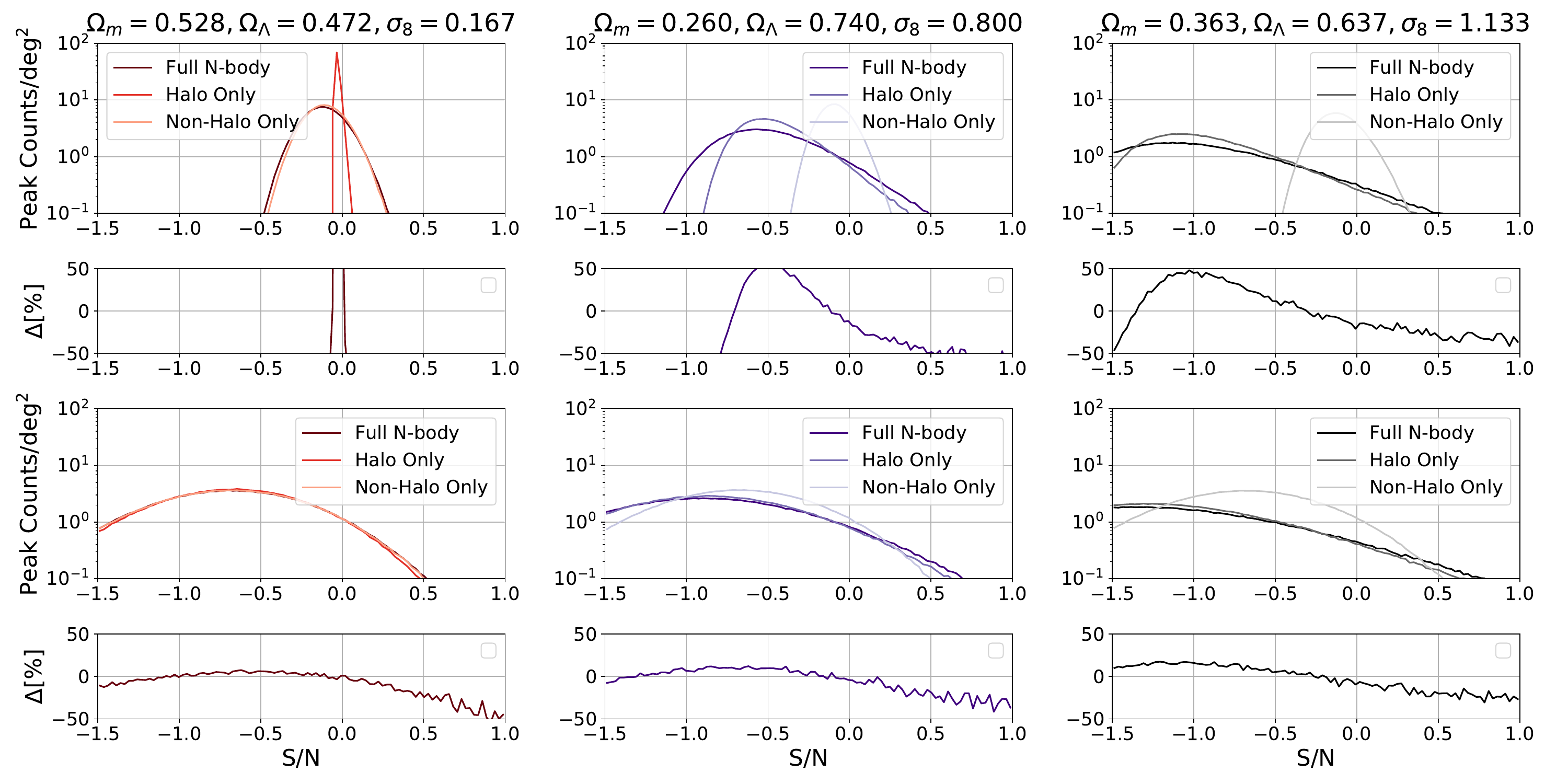}
\caption{\label{fig:minimahist} Mean minima counts for three different cosmologies: low-$\Sigma_{8}$ (left), fiducial (center), and high-$\Sigma_{8}$ (right). As in Fig.~\ref{fig:peakhist}, the top panel shows mean minima counts from smoothed WL maps, while the bottom panel shows peak counts after both smoothing and noise were added. The histograms are based on 100 bins(S/N=[$-\infty,-1.5, ..1.0,+\infty$]. 
The second and fourth row of panels show the fractional differences between full N-body and halo-only maps ($\Delta=(N_{\rm halo} - N_{\rm full})/N_{\rm full}$).}
\end{figure*}

Ref. \cite{Coulton2020} recently investigated WL minima as an additional statistic for constraining cosmological parameters.  
Minima are defined analogous to peaks, except they represent pixels whose convergence is smaller than those in the 8 adjacent pixels.
Ref. \cite{Coulton2020} found that baryonic effects result in only $\approx0.5\sigma$ biases in the values of cosmological parameters inferred from these minima, in comparison to $\approx4\sigma$ biases when WL peaks are used instead. Motivated by these findings, we look at the minima counts in our full N-body, halo-only and non-halo-only maps. Histograms of these minima are shown in Fig.~\ref{fig:minimahist}. They reveal relationships similar to those in the peak histograms: halo-only maps do not explain all the negative minima. The median differences between halo-only and full N-body minima counts for low-$\Sigma_{8}$, fiducial and high-$\Sigma_{8}$ cosmologies are $\approx 9\%$, $6\%$ and $13\%$, all similar to or notably higher than the residuals for peak counts. Halo-based models are able to predict minima counts worse than peaks. This further explains the lack of sensitivity of minima counts to baryons, because WL minimas are associated with the material residing outside halos more than peaks (see also \cite{Maturi2010}). Therefore, in order to take advantage of this new statistic, modified models, beyond halos, need to be implemented.

\subsection{Limitations and Future Work}

There are several limitations and idealizations in our analysis, which we only briefly highlight here.  First, by using simulations alone in our analysis, we do not account for any systematic uncertainties that may be present in weak lensing data. Additionally, our simulated maps have all source galaxies placed at a single redshift and as we have seen in \S~\ref{discussion}, peak count histograms are strongly redshift-dependent. We also do not account for baryonic effects in our study, which have shown to affect the high S/N peaks \cite{Yang2013} and have a biasing effect on cosmological constraints from peaks more generally \cite{Coulton2020,LuHaiman2021}. In order to implement and improve models for peak counts, these effects need to be addressed in future studies.    

Nevertheless, the main conclusion of our study, namely that material outside halos must be incorporated into any model of lensing peak counts to reach percent-level accuracy, remains robust.

\section{Summary and Conclusion}

In this work we have shown that ``perfect'' halo-based models, which include realistic positions and shapes of halos, i.e. directly taken from N-body simulations, still do not account for peak counts at the percent-level accuracy, needed for precision cosmology with next-generation weak lensing surveys. Using halo-only models for peak counts results in biased cosmological constraints for surveys $\gtrapprox 100$ deg$^{2}$.   We conclude that fast, halo-based models for peak counts must incorporate the impact of mass outside and far away from dark matter halos, before they can be applied to these data for cosmological parameter inference. 

By separating different effects in the halo-only weak lensing maps, we were able to show that halo clustering has the most significant effect on predicting peak counts, whereas halo shapes (replacing simulated halos by halos with NFW profiles and fixed concentration parameters) have a smaller effect, near the peak of the peak-count distribution. Therefore, incorporating realistic clustering into current halo models should noticeably improve their predictions.  On the other hand, for high positive peaks, halo clustering and halo shapes have comparable but opposing effects, resulting in a near cancellation.  This, somewhat coincidentally, makes predictions from the halo-based CAMELUS model accurate in this regime. We caution that the behavior of this cancellation, such as its dependence on cosmology, remains to be investigated.

We have also looked at the material surrounding halos and learned that mass residing in the large-scale structures far from halos (beyond 15 times their virial radii) is still important for predicting peak counts accurately, especially for negative peaks. Including the material only on the periphery of halos does not alleviate the differences in peak counts to the percent level. Therefore, incorporating filaments and voids, which are located far from halos, will likely improve current halo-based models for peak counts. 

\section{Acknowledgements}

We thank Peter Behroozi for helpful discussions about \texttt{Rockstar} and the anonymous referee for useful comments on our manuscript. This work was supported by NASA ATP grant 80NSSC18K1093. We acknowledge the use of the NSF XSEDE facility for the simulations and data analysis in this study. This research made use of open source \texttt{Python} packages \texttt{NumPy} \cite{numpy}, \texttt{SciPy} \cite{2020SciPy} and \texttt{Matplotlib} \cite{matplotlib}. This research made use of \texttt{Astropy}, a community-developed core Python package for Astronomy \cite{astropy:2013, astropy:2018}
\bibliography{apssamp}
\end{document}